# Development of a 3-Dimensional Dosimetry System for Leksell Gamma Knife-Perfexion

**KyoungJun Yoon, JungWon Kwak, DoHeui Lee, ByungChul Cho, SangWook Lee, SeungDo Ahn**

*Department of Radiation Oncology, Asan Medical Center, University of Ulsan College of Medicine, Seoul Korea 138-736*

The purpose of our study is to develop a new, 3-dimensional dosimetry system to verify the accuracy of dose deliveries in Leksell Gamma Knife-Perfexion™ (LGKP) (Elekta, Norcross, GA, USA). The instrument consists of a moving head phantom, an embedded thin active layer and a CCD camera system and was designed to be mounted to LGKP. As an active material concentrically located in the hemispheric head phantom, we choose Gafchromic EBT3 films and Gd2O2S;Tb phosphor sheets for dosimetric measurements. Also, to compensate the lack of backscatter, we located a 1 cm thick PMMA plate downstream of the active layer. The PMMA plate was transparent for scintillation lights to reach the CCD with 1200×1200 pixels by a 5.2 pitch. Using this system, three hundred images by a 0.2 mm slice gap were acquired under each of three collimator setups, i.e. 4 mm, 8 mm, and 16 mm, respectively. The 2D projected images taken by CCD camera were compared with the dose distributions measured by EBT3 films in the same conditions. All 2D distributions were normalized by the maximum values derived by fitting peaks for each collimator setup. The FWHM differences of 2D profiles between of CCD images and of film doses were measured to be less than 0.3 mm. The scanning task for whole peak regions took less than three minutes with the new instrument. It could be utilized as a QA tool for the Gamma Knife radiosurgery system instead of film dosimetry, the use of which requires much more time and resources.

Email: jwkwak0301@gmail.com

Fax: +82-2-3010-5731

## I. INTRODUCTION

Radiosurgery is one of the most relevant techniques for radiation treatments of small brain tumors. The Gamma Knife system is a Co-60 based medical device for treatment target irradiations in the radiosurgery [1-4].

Leksell Gamma Knife-Perfexion™ (LGKP) (Elekta, Norcross, GA) is equipped with three, different collimator sizes; 4, 8 and 16 mm in diameter to focus the gamma rays on a common spot called the field isocenter. The most critical change in the Perfexion system is the new collimation system which is fully automated and built in the Gamma Knife system.

Since the gamma rays are emitted from 192 Co-60 sources through collimators which are grouped in eight sectors individually operated, the beam shapes can be dynamically changed according to the sector cone configurations [5]. There is a theoretical possibility of choosing from 65,000 combinations of highly focused radiation beams. Radiosurgery characterized by a single or high-dose fractionation, requires regular and careful quality assurance [6, 7].

Checking the integrity of collimator systems is one of the important quality assurance tasks required in order to ensure the accuracy of the dose delivery. Although a simple visual inspection was sufficient to verify the geometrical configuration of the collimators in the previous LGKP system, there is no simple way to independently evaluate the built-in collimator of the LGKP[8].

The conventional method of exposing film at the isocenter provides a composite dose image which has been successfully used for checking the coincidence between the radiation and the mechanical isocenter and also for dose-rate verification [9, 10]. However, it is difficult to use the film in terms of the integrity of each individual source and the corresponding collimator configuration system. The purpose of our study is to develop a method of verifying the dosimetric accuracy for the collimation configurations using a fluorescent screen with a CCD camera in LGKP.

## II. EXPERIMENTS

We designed and fabricated a new detector system to verify the 2D and 3D dose distributions from the irradiation fields under various collimation sector configurations in LGKP.

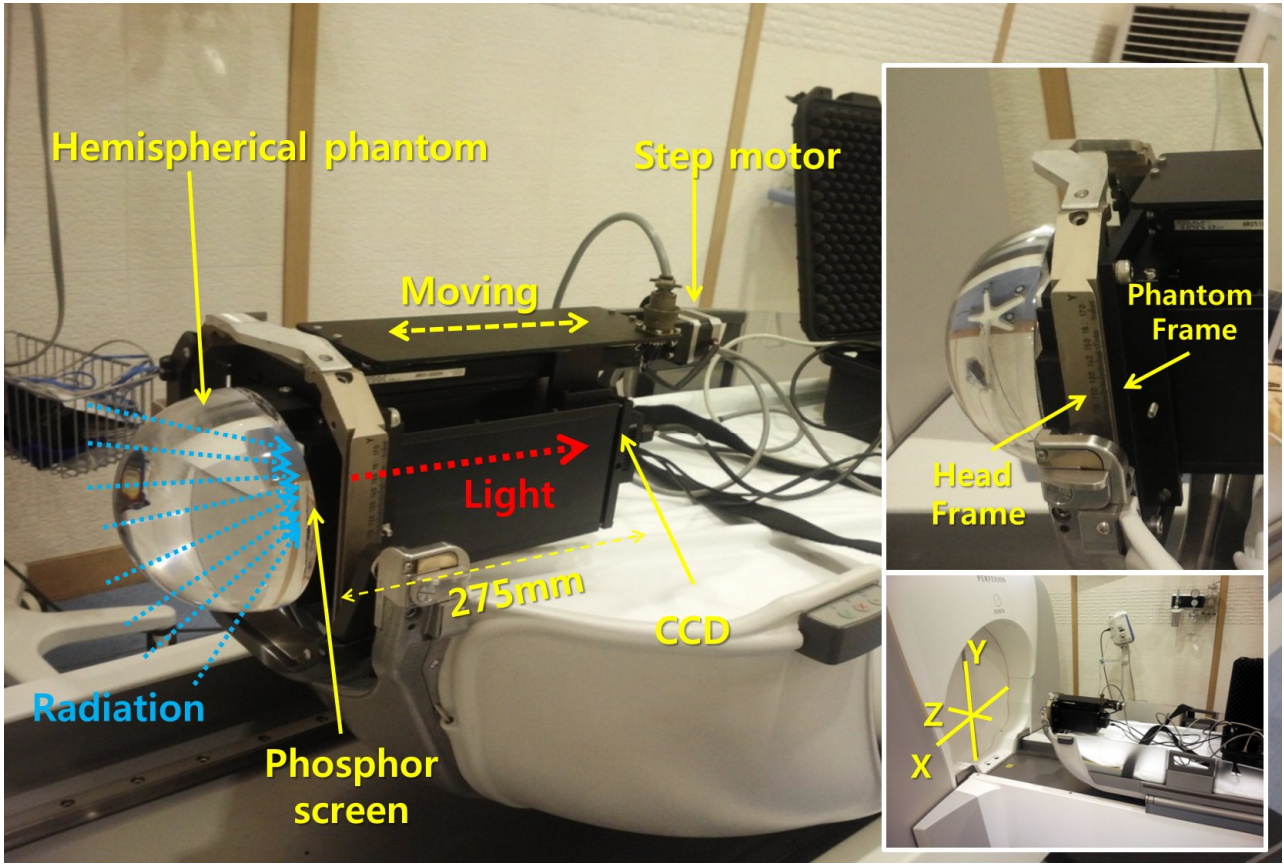

Fig. 1. The new detector devices installed in the LGKP for experimental setup.

Fig. 1 shows the devices and the experimental setup. The detector system could be mounted to LGKP combining the gamma knife head frame to a phantom frame. Beginning measurements, the home position of the hemispheric head Phantom center was calibrated to the beam center. Then, the phosphor screen of the system was positioned at the beam field center position. The phantom part including the active layer and CCD camera could move ± 30 mm in superior-inferior direction from the home position ( Z = 0 mm ). The mobile CCD-based detection system was allowed simultaneous imaging of gamma ray paths from the individual sources. The scan images were acquired under each collimator size with the 0.5 sec exposure time taken in the single frame acquisition. The acquirement of 300 image frames with a 0.2 mm slice gap could cover enough ranges of the single shot under any collimator configuration in LGKP. To evaluate the dosimetric characteristics of the fluorescent screen, the comparison with the Gafchromic film (EBT3, International Specialty Products, Wayne, NJ, USA) measurements was performed. The material composition of the employed phosphor thin layer (Gd2O2S:Tb, Fujifilm Corp, Japan) has been demonstrated as an excellent material for ionizing radiation imaging and dosimetry [11,12].

The main concern for the phosphor dosimetry was related with image processing, since there were lots of scattered spots in CCD images. The scattered spots occupied in single pixels originated from the direct interaction of the scattered radiations from the head phantom with CCD. An image correction process using a median filter in imageJ (http://rsb.info.nih.gov/ij/index.html) program was employed to remove these unnecessary spots. In the filtering process, the pixel was replaced by the median of pixel values within a distance of radius (2 pixels).

CCD camera (VA-2MG-42, 1200×1200 pixels, 5.2 ×5.2 pixel size, Syncron Corp, Korea), Lens (FV1520, Syncron Corp, Korea), Hemispherical phantom (HP) (diameter 16 cm, PMMA ), system controller board, and step motor to move the system and a control computer were installed in this system.

The head frame of the LGKP was attached to the phantom frame. The home position of HP center was calibrated to the isocenter with its axis aligned along the longitudinal (Z)-axis of the couch.

The images which could be scanned at various programmable positions were saved in 12bit raw data formats in an image control unit.

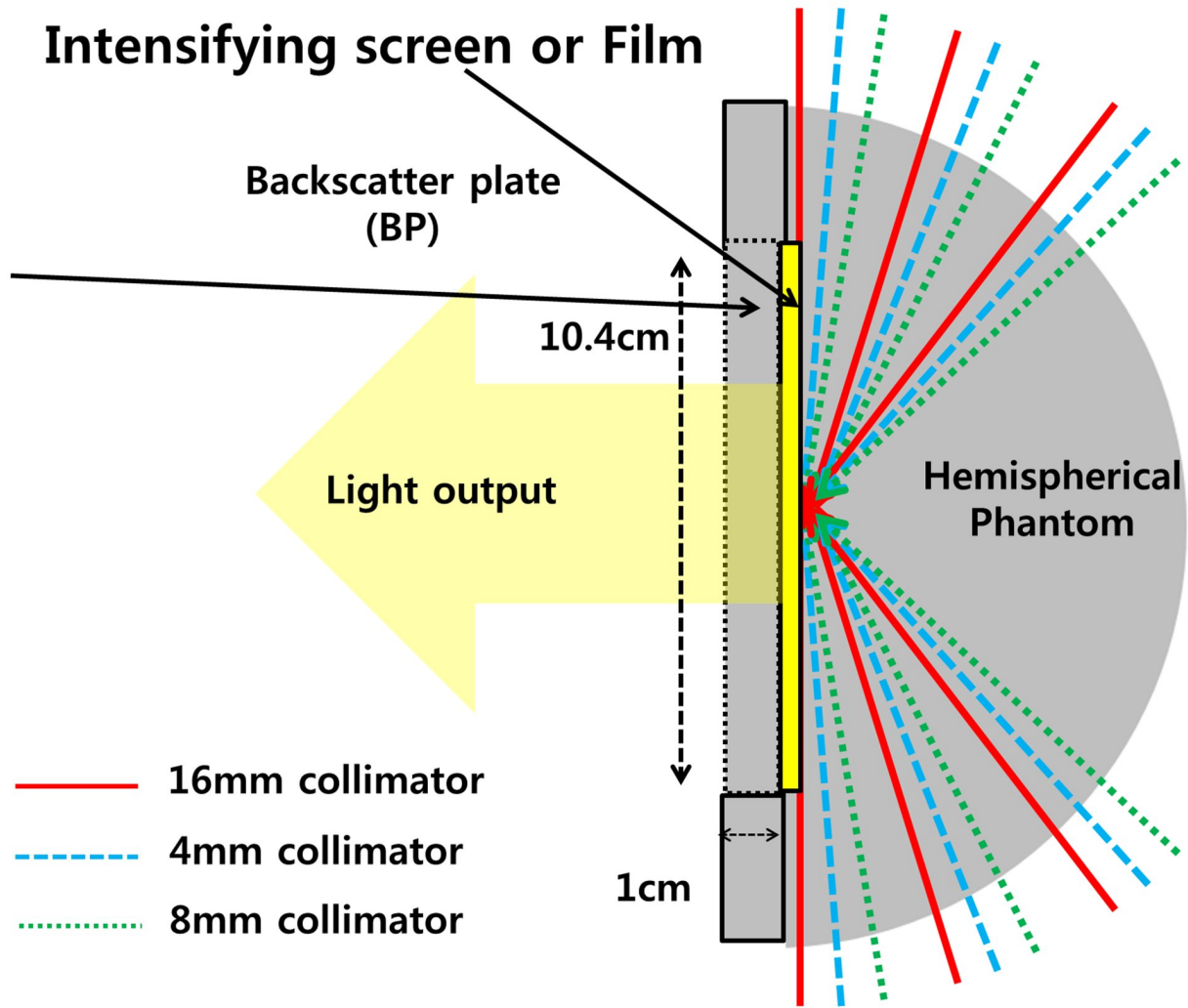

Fig. 2. The head phantom with an active layer; a scintillator screen or EBT3 film: The intensifying screen was required to be located between the HP and the BP to compensate the lack of back scattering.

As shown in Fig 2, the backscatter plate (BP) made of a 1 cm thick PMMA slab was combined with the HP for backscatter compensating. The PMMA material for BP was known to be transparent at the 545 nm wave length of scintillation lights to be able to reach the CCD[13].

Table 1. Output factors for various collimator sizes in Gamma Knife system and experimental beam delivery configurations.

| Collimator | Effective output factors | Prescription | Beam-on time |
|---|---|---|---|
| 4 mm | 0.805 | 2.0Gy@50% | 111 s |
| 8 mm | 0.924 | 2.0Gy@50% | 97.2 s |
| 16 mm | 1 | 2.0Gy@50% | 88.8 s |
| Treatment Dose rate 2.680 Gy/min | | | |

As shown in Table 1, the film irradiation times were individually set to be 111 s, 97.2 s, and 88.8 s for each collimation size for 2 Gy prescription at the 50% isodose level using the Leksell GammaPlan@9.0.0. After the irradiated EBT3 films were scanned using a scanner (Epson 10000XL, NISCA Corporation Inc., 72 dpi, 600×600 pixels.), the dose conversion processes were performed with FilmQA program (Ashland Inc, Bridgewater, NJ, USA). The 2D projected images from the 3D CCD image sets were analyzed in comparison with the film dose distributions.

## III. RESULTS AND DISCUSSION

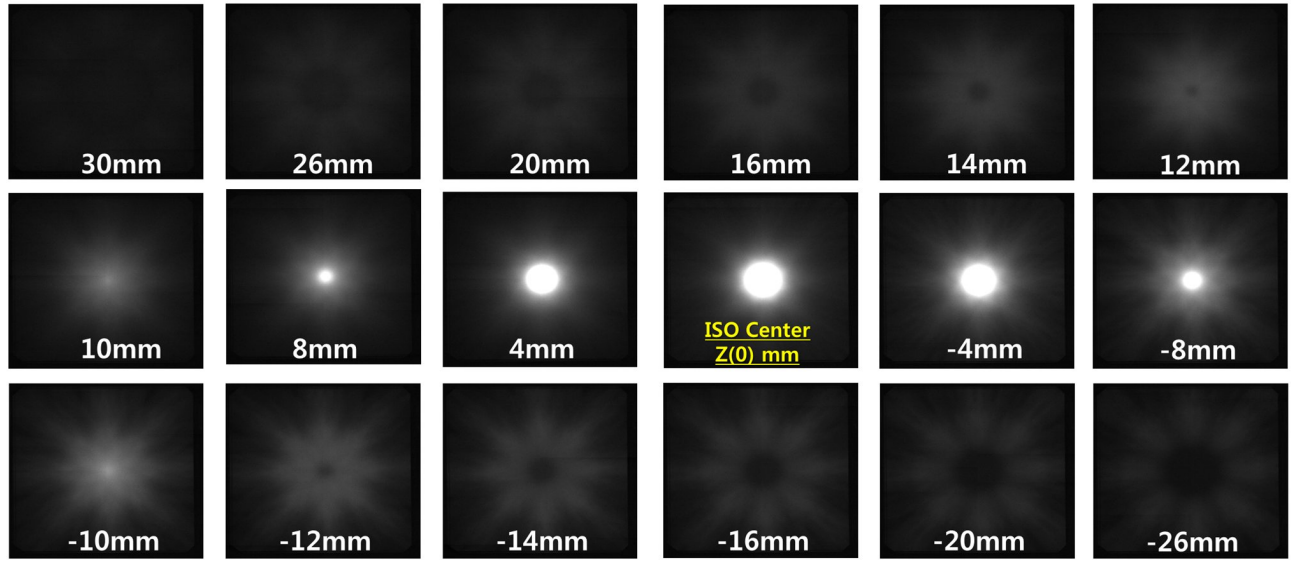

Fig. 3. Selected axial CCD images taken under 16 mm collimation. Original data consists of three hundred images by 0.2 mm slice gap.

Fig. 3 shows the axial CCD images acquired under 16 mm collimation configuration. We could recognize the 192 beam tracks from collimated sources grouped in 8 sectors. The window level of the CCD images was adjusted to enhance the lower intensity level using imageJ program. The axial level with the highest enhanced images was interpreted as the home position of the detector ($Z = 0$ mm).

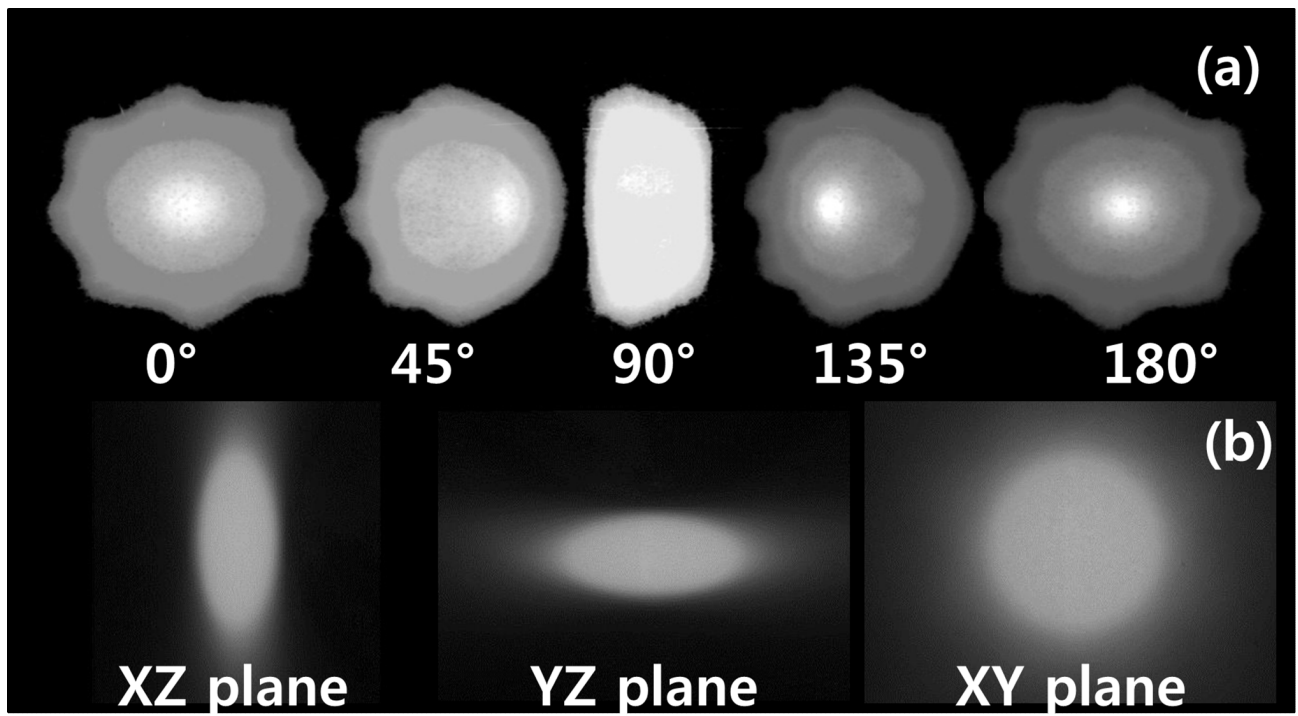

Fig. 4. (a) The 3D image taken under 16 mm collimation which was reconstructed by combining 2D axial images. (b) 2D sliced images in sagittal, coronal and axial views derived from the reconstructed 3D image at home position of detectors.

The 3D scan images were reconstructed with the 2D axial images acquired with 16 mm collimator using the ImageJ program. The 2D images shown in Fig. 4(a) were generated by projecting the reconstructed 3D image set in various azimuthal angles; 0˚, 45˚, 90˚, 135˚, and 180˚. As shown in Fig 4(b), 2D sliced images for 16 mm collimation were shown in sagittal, coronal and axial views.

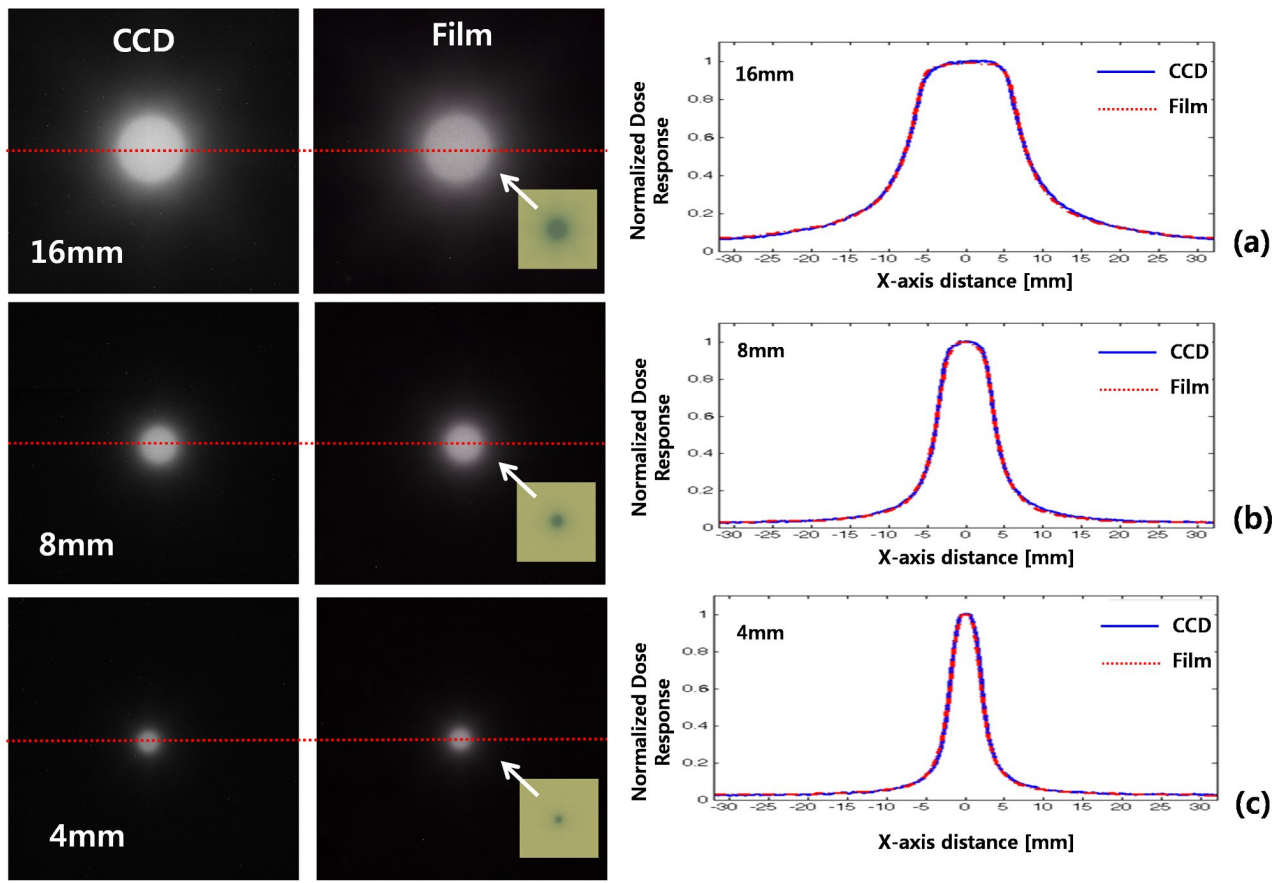

Fig. 5. Comparison between the CCD images and the film measurement images. The dose profile curves comparison of normalized film and the CCD dose response with 4 mm, 8 mm, and 16 mm collimators. The solid line shows the profile of the CCD system. The dotted line shows the profile of the film. (a) is the 16 mm collimator profile, (b) is the 8 mm profile, and (c) is the 4 mm profile.

In Fig. 5, the comparisons between the CCD images and the film dose measurements were performed in 2D image maps (Right) and in the line profiles of the X-axis (Left) for the three collimation conditions. Each profile was normalized by the maximum values derived with fitting peak regions. Two aspects ; FWHM values of peaks and profile shapes in the dose tail regions were considered. FWHM differences between the measured from Film and CCD images values were 0.1 mm for 4mm collimator, -0.3 mm for 8 mm collimator, and 0.2 mm for 16 mm collimator. For the tail shapes for all conditions, any systematical inconsistence was not recognized. According to these results, we confirmed that the data of the CCD-based QA system were in good agreement with the dose measurements of film as listed in Table 2, while the acquisition time of the 300 images using this equipment took less than three minutes.

Table 2. Comparison of FWHM between the film measurement images and the CCD images .

| Collimator | Film measured FWHM (mm) | CCD measured FWHM (mm) | Absolute difference |
| --- | --- | --- | --- |
| 4 mm | 6 | 6.1 | 0.1 |
| 8 mm | 10.8 | 10.5 | -0.3 |
| 16 mm | 21.3 | 21.5 | 0.2 |

## IV. CONCLUSION

We developed a CCD camera–based 3D dosimetry system for quality assurance of the Gamma Knife. The FWHM differences of the profile curves between the Film measured and the CCD values were less than 0.3 mm. The CCD images were in good agreement in profile shape. As a result, we confirmed that the system could be a useful constitute of film dosimetry for QA tasks of the Gamma Knife.

## ACKNOWLEDGEMENT

This research was supported by the Basic Science Research Program through the National Research Foundation of Korea (NRF) funded by the Ministry of Education, Science and technology(2014R1A1A2058154).